# Agentic metasurface design with self-correcting language-model systems

Bei Wu[1,2,3,4], Bo Xiong[1,2,3,4], Haiyao Luo[1,2,3,4], Yaqi Li[1,2,3,4], Li Zhang[1,2,3,4], Qiaolu Chen[1,2,3,4,5], Hongsheng Chen[1,2,3,4*] and Yihao Yang[1,2,3,4*]

*[1] State Key Laboratory of Extreme Photonics and Instrumentation, ZJU-Hangzhou Global Scientific and Technological Innovation Center, Zhejiang University, Hangzhou 310027, China.*
*[2] International Joint Innovation Center, The Electromagnetics Academy at Zhejiang University, Zhejiang University, Haining 314400, China.*
*[3] Key Lab. of Advanced Micro/Nano Electronic Devices & Smart Systems of Zhejiang, Jinhua Institute of Zhejiang University, Zhejiang University, Jinhua 321099, China.*
*[4] Shaoxing Institute of Zhejiang University, Zhejiang University, Shaoxing 312000, China.*
*[5] Laboratory of Wave Engineering, Ecole Polytechnique Fédérale de Lausanne (EPFL), Lausanne, Switzerland.*

*[*] Corresponding authors: hansomchen@zju.edu.cn (H. Chen); yangyihao@zju.edu.cn (Y. Yang)*

## Abstract

Automated metasurface design is increasingly important, and recent advances in language-model systems are opening a route toward agentic optical design. Yet modern metasurface applications, from metalenses and holography to optical computing, require long design chains spanning modeling, simulation, coding, optimization and evaluation. These chains are error-prone, whereas existing language-model-based metasurface tools remain largely limited to simple objectives, predefined pipelines or language-to-layout generation. Here we introduce MetaDesigner, a self-correcting language-model system for agentic metasurface design. From a natural-language optical objective, MetaDesigner plans the design route, retrieves domain knowledge, invokes simulation and optimization tools, generates missing tool code and identifies errors through a dedicated Verifier. We demonstrate three tasks of increasing complexity: an RGB metalens with three independent focal spots, a six-plane full-color hologram with an average structural similarity index measure (SSIM) of 0.97, and an optoelectronic hybrid neural network for image style transfer. These tasks require 74, 136 and 90 reasoning steps, respectively, and the system self-corrects errors in frequency mapping, numerical aperture estimation, network-parameter counting and loss-function description. These results establish MetaDesigner as a self-correcting route to agentic metasurface design, where language-model systems can not only execute optical design tasks but also extend, inspect and repair the design process itself.

## Introduction

Optical metasurfaces[1-3] have reshaped optical design by replacing bulky refractive and diffractive components[4] with ultrathin arrays of engineered nanostructures. By tailoring phase[5,6], polarization[7,8], amplitude and momentum[9,10] at the subwavelength scale, they have enabled compact beam shapers[11], metalenses[12,13], holograms[14-16] and diffractive computing systems[17-19]. This versatility arises from an exceptionally large design space, in which material choice, unit-cell geometry and spatial arrangement can all serve as degrees of freedom. Over the past decade, adjoint optimization, topology optimization[20], evolutionary search[21] and deep learning[22,23] have made parts of this space increasingly searchable. Yet these methods typically operate only after the essential design problem has been formulated by human experts. The designer must still decide which physics to model, which variables to optimize, which objective to impose, which solver to use and how to interpret the result. Thus, although modern inverse design has made metasurface optimization increasingly automated, the construction of the design process itself remains largely human-driven.

Language-model systems offer a route from automated optimization toward agentic metasurface design. By combining natural-language reasoning, domain-knowledge retrieval, code generation, tool use and iterative revision, they can in principle operate at the level where an optical intention is converted into an executable design process[24-29]. This level is particularly important for advanced metasurfaces. A functional device is not produced by optimization alone, but by a coupled chain of decisions involving physical modeling, electromagnetic simulation, data preparation, numerical propagation, loss-function design, code implementation, performance evaluation and technical interpretation. Recent agentic frameworks have begun to show that language-described photonic goals can be converted into device layouts[30,31]. However, as optical tasks become more complex, this chain becomes increasingly fragile. A wrong wavelength convention, an inappropriate propagation model, a hidden coding error or an overinterpreted metric can pass through optimization and yield a result that appears plausible but is physically unreliable. Agentic metasurface design therefore requires more than the ability to execute tasks: the system must also be able to inspect, verify and repair its own reasoning and computation.

Here we introduce MetaDesigner, a self-correcting language-model system for agentic metasurface design. Unlike previous metasurface agents that mainly translate language-described goals into device layouts within predefined pipelines, MetaDesigner plans, executes and verifies the design process itself. Starting from a natural-language optical objective, it decomposes the task into long-horizon design steps, retrieves domain knowledge, invokes simulation and optimization tools, generates missing tool code when needed and produces technical reports. Reliability is built into this process through a dedicated Verifier, which cross-checks reasoning trajectories, generated code, numerical results and reports, allowing errors to be detected and corrected within the design loop. We demonstrate MetaDesigner on three tasks of increasing complexity: an RGB metalens with three separate focal spots, a six-plane full-color hologram with an average SSIM of 0.97, and an optoelectronic hybrid neural network for image style transfer. These tasks require 74, 136 and 90 reasoning steps, respectively, spanning task interpretation, physical modeling, code implementation, numerical optimization, evaluation and report generation. Across these demonstrations, the Verifier successfully catches and corrects errors in frequency mapping, numerical aperture estimation, network-parameter counting and loss-function description, highlighting that MetaDesigner is not only autonomous long-horizon metasurface design but also actively monitors and repairs its own design loop.

## Results

### A self-correcting language-model system

MetaDesigner is a self-correcting multi-agent system that completes metasurface design tasks from user-provided natural-language prompts. Given an optical objective, MetaDesigner orchestrates multiple specialized sub-agents to accomplish tasks collaboratively, generating runnable code, optimized network models, metasurface structures, and numerical results that are stored in a local file system (LFS). It also records the reasoning trajectory for inspection and generates Markdown-format technical reports that document the background research, design rationale, implementation details, optimization results, and performance evaluation. Rather than treating LLMs as passive text generators, MetaDesigner structures its reasoning capability into a closed-loop design cycle encompassing planning, execution, verification, and correction[32].

The language-model system comprises five specialized agents (Fig. 1a): the Solver, Verifier, Researcher, Optimizer, and Programmer, along with three specialized tools: electromagnetic simulation, retrieval-augmented generation (RAG), and file management. The Solver, Verifier, Researcher, and Optimizer use DeepSeek-V3[33] as the base model, which provides stable long-horizon reasoning behavior in our implementation. The Programmer uses Claude Sonnet 4.5 as the base model and is responsible for missing tool code generation. The core of MetaDesigner is the collaboration between the Solver and Verifier. As depicted in Fig. 1b, the Solver interprets user prompts, decomposes the design objectives into executable sub-tasks, and invokes appropriate sub-agents or tools to execute them. The Verifier examines the Solver's reasoning trajectory and provides corrective feedback on unsupported assumptions and hallucinations. It inspects generated technical reports, code, and numerical results, and re-invokes sub-agents or tools to cross-check key claims and outputs, enabling evidence-based verification instead of relying solely on linguistic critique. The Solver adopts the feedback after justifying it, thereby refining its reasoning without blindly obeying the Verifier.

Each sub-agent and tool shown in Fig. 1c provides distinct specialized capabilities. Specifically, the Researcher retrieves scientific literature through arXiv and searches online information through TavilyClient; the electromagnetic simulation tool models meta-atom scattering responses using Computer Simulation Technology (CST); and the Optimizer designs metasurface phase profiles through physics-based diffraction and coherence[34]. When a required capability is not available in the existing tool library, the LLM agents invoke the Programmer to generate tool code to fill the gap. To improve the accuracy of the generated code, users can construct a RAG database with relevant technical papers or webpages[35], enabling the LLM agents to draw on task-specific background knowledge while requiring users only to supply materials rather than master technical details. To avoid unintended overwriting of local files, MetaDesigner stores generated files in a virtual file system (VFS) and uses a conflict-checking file management tool to coordinate transfers between the VFS and the LFS.

To support long-horizon reasoning, the language-model system employs a structured memory management module. The Solver and Verifier share persistent memory stored in a PostgreSQL database, enabling direct access to each other's reasoning trajectories and verification processes. In contrast, specialized sub-agents use isolated in-memory storage, delivering only synthesized conclusions rather than full reasoning trajectories. Beyond shared memory, the LLM agents communicate through autonomous dialogues. For example, the Solver assigns literature research sub-tasks to the Researcher using natural-language prompts, such as: "What is an optoelectronic hybrid neural network for image style transfer?" After completing the search, the Researcher returns a synthesized conclusion of the retrieved knowledge to the Solver.

We evaluate MetaDesigner on a hierarchy of optical metasurface design tasks — from RGB metalens design[12,13] and computer-generated holography (CGH)[14-16] to optical generative modeling[36,37] — which progress from multi-wavelength point focusing (Fig. 1d) to multi-plane image reconstruction (Fig. 1e) and finally image style transfer (Fig. 1f).

**Autonomous RGB metalens design**

In the first benchmark task, MetaDesigner is required to design an RGB metalens that focuses normally incident red, green, and blue light onto three distinct positions on a focal plane located 80 μm behind the metalens. The operating frequencies of the red, green, and blue components are 480, 560, and 640 THz, corresponding to wavelengths of approximately 625, 536, and 469 nm, respectively. Their target focal positions are defined as $(x, y) = (0, -64\Delta)$, $(0, 0)$, and $(0, 64\Delta)$, where $\Delta = 160$ nm is the unit-cell period. The design objective is therefore to generate a metalens phase profile composed of 512 × 512 meta-atoms that produces the target optical field distribution.

Figure 2a shows the agentic workflow for RGB metalens design. The Solver analyzes the design requirements, formulates an executable workflow, invokes appropriate tools from the library, and generates tool code when existing tools are insufficient. As illustrated in Fig. 2b, the available tools include TavilyClient for online information search, arxiv for scientific literature retrieval, CST for simulating meta-atom scattering responses, Claude Sonnet 4.5 for missing tool code generation, RAG for retrieval-augmented generation from curated technical knowledge, and a file management tool for transferring files between the VFS and the LFS. As a proof of concept, the meta-atom consists of a $TiO_2$ nanopillar on a $SiO_2$ substrate, with the unit-cell period and nanopillar radius fixed at 160 nm and 50 nm, respectively. The transmitted phase is tuned by varying the nanopillar height from 200 to 800 nm, as shown in Fig. 2c. Using the phase at 480 THz as a reference, the CST-based electromagnetic simulation tool extracts empirical linear relationships among the phase shifts at different frequencies, allowing the Optimizer to estimate multi-frequency phase responses directly from the reference phase.

As shown in Fig. 2d, MetaDesigner completes the RGB metalens design through 74 reasoning steps, taking 16 min 14 s and consuming 1.89 million tokens. The reasoning trajectory proceeds through five stages. In stage S1, the Researcher acquires domain knowledge on RGB metalens principles, multi-wavelength phase control, chromatic-aberration correction, diffraction-limited focusing, and metalens dispersion engineering, which informs the subsequent workflow. In stage S2, the Optimizer constructs the computational design pipeline, including target intensity generation, optical propagation modeling based on the angular spectrum method (ASM)[38], metalens phase optimization by back-propagating the loss between the simulated and target optical intensities, and quantitative performance evaluation.

Figure 2e shows that the optimized metalens generates three well-separated focal spots. The intensity profiles along the *y*-axis further confirm that the red, green, and blue components are focused at their target positions (Fig. 2f). The focusing efficiency, defined as the ratio of optical power within the focal spot to the incident optical power, averages 20% across the three frequencies (Fig. 2g). This trade-off is expected for a proof-of-concept multi-functional RGB metalens, where a single metalens must simultaneously direct three wavelengths to three independent focal spots. Such multi-wavelength and multi-functional operation generally introduces efficiency trade-offs compared with single-wavelength or single-function metalenses[39,40]. In our implementation, the efficiency is further constrained by our deliberately simplified meta-atom geometry (varying only the nanopillar height), and can be improved by expanding the design space to include more geometric or material parameters[41-43].

In stage S3, the Solver consolidates the optimization results and compiles them into structured technical reports, as detailed in Supplementary Note 6. In stage S4, the Verifier cross-checks the reasoning trajectory, generated code, optimization results, and technical reports. This process identifies several substantive errors, including color-frequency mapping errors, numerical aperture calculation errors, overstatement that the focal spots exceed the diffraction limit, and inconsistent descriptions of the same viewpoint across different technical reports. In stage S5, the Solver corrects these errors sequentially through the verification-and-correction loop, as detailed in Supplementary Notes 7-8. These results demonstrate that MetaDesigner can autonomously complete a full metalens design cycle, spanning task interpretation, knowledge retrieval, electromagnetic simulation, differentiable optical optimization, evidence-based verification, and report generation. We next increase the task complexity from multi-wavelength point focusing to multi-plane image reconstruction.

**Multi-plane computer-generated holography**

Building on the wavefront engineering capability demonstrated in the RGB metalens task, we next challenge MetaDesigner with multi-plane CGH. As illustrated in Fig. 3a, the design objective is to simultaneously reconstruct full-color holographic images at six image planes, $z_i$ = 51, 52, 53, 54, 55, and 56 μm. To achieve this, we use a transmissive metasurface composed of 512 × 512 meta-atoms to phase-modulate normally incident red, green, and blue light at 480, 560, and 640 THz, respectively.

MetaDesigner completes the multi-plane CGH task through 136 reasoning steps, taking 47 min 27 s and consuming 5.62 million tokens (Fig. 3b). The reasoning trajectory proceeds through four stages. In stage S1, the Researcher acquires task-relevant knowledge on metasurface-based CGH, including RGB phase control, optical propagation modeling, optimization methods for CGH, full-color holography advances, and challenges such as dispersion, cross-talk and limited degrees of freedom, with the retrieved knowledge summarized in Fig. 3c. The Solver then converts the user-provided RGB-D dataset, as shown in Fig. 3d, into tensor-format data that can be directly processed by the Optimizer. Specifically, the depth map is quantized into six depth levels, assigning RGB image content to the corresponding image planes and yielding target RGB intensity distributions $I(x, y, z_i)$.

In stage S2, the Optimizer constructs the computational design pipeline for multi-plane CGH, including depth-resolved optical propagation, metasurface phase optimization, and quantitative evaluation of simultaneous image reconstruction. Specifically, it seeks a metasurface phase profile $\Phi(x, y)$ whose propagated intensity distributions match the target ones at all six image planes. The optimization problem is expressed as $\mathrm{argmin}_{\Phi} \sum_{i=1}^{6} \left\| \left| Prop_{z_i}(e^{i\Phi(x,y)}) \right|^2 - I(x, y, z_i) \right\|_2^2$, where $Prop_{z_i}$ denotes the free-space propagation operator based on the ASM, as depicted in Fig. 3e. The optimized phase profile at 480 THz is presented in Fig. 3f. The phase profiles at 560 THz and 640 THz can be derived from this reference phase using the empirical linear relationships obtained from the meta-atom response model. As shown in Fig. 3g, the reconstructed multi-plane holographic images show good agreement with the target. The Programmer further generates evaluation code for peak signal-to-noise ratio (PSNR) and SSIM (Fig. 3h). Averaged over the multi-plane holograms, PSNR and SSIM, computed over the masked target regions, reach 37.85 dB and 0.97, respectively, demonstrating high-fidelity reconstruction.

In stage S3, the Solver analyzes the optimization results and compiles them into structured technical reports, as detailed in Supplementary Note 11. In stage S4, the Verifier cross-checks the reasoning trajectory, generated code, dataset, optimization results, and technical reports. This process identifies several substantive errors. First, the dataset creation process incorrectly concatenates the image channels as [G, B, R] rather than [R, G, B], causing the target hologram channels to be misaligned with the propagation frequencies. Second, the initial SSIM and PSNR are therefore obtained from a channel-mismatched evaluation and are revised after correcting the channel order, yielding an average SSIM improvement from 0.9633 to 0.9705 and an average PSNR improvement from 36.75 dB to 37.85 dB. Third, the generated technical report incorrectly describes the reconstruction images as a single file named Part0-5.png; in fact, the outputs are saved as six separate images, Part0.png to Part5.png. The Solver subsequently corrects this error, with the detected error and correction summarized in Fig. 3i. Details of the verification-and-correction loop are provided in Supplementary Notes 12-13. These results demonstrate that MetaDesigner can autonomously complete multi-plane full-color holographic reconstruction from fixed target intensity distributions. We next move beyond fixed-target reconstruction and construct an optical generative model for image style transfer.

**Optical generative modeling for image style transfer**

In this task, MetaDesigner is employed to design an optoelectronic hybrid neural network[44-46] for image style transfer. The network consists of a shallow digital encoder and a phase-modulated metasurface decoder. Given a content image and a target style image, the objective is to generate an image that preserves the structural features of the content image and the texture features of the style image. In the designed optoelectronic hybrid neural network, the digital encoder maps each input content image into the RGB amplitude profiles of the incident optical field. The encoded optical field is then phase-modulated by the metasurface and propagates through free space, with the metasurface and propagation jointly functioning as an optical decoder. The resulting optical intensity distribution is recorded by an image sensor placed 50 μm behind the metasurface, forming the style-transferred output image.

MetaDesigner completes the optical generative modeling task through 90 reasoning steps, taking 27 min 57 s and consuming 2.53 million tokens (Fig. 4a). The reasoning trajectory proceeds through four stages. In stage S1, the Researcher investigates optoelectronic hybrid neural networks for image style transfer, covering optical neural networks[47], ASM-based optical propagation, CGH, and neural style transfer. In stage S2, the Optimizer constructs the computational design pipeline for optoelectronic style transfer, including dataset preparation and validation, encoder network implementation, optical propagation modeling, VGG19-based perceptual loss implementation, and joint digital-optical optimization.

The architecture of the designed optoelectronic hybrid neural network is shown in Fig. 4b. During training, the digital encoder and the metasurface phase profile are jointly optimized so that the generated images match the content images in structure and the style images in texture. Guided by the RAG database, the Programmer implements the perceptual loss functions using the convolutional feature extractor of VGG19[48] pretrained on ImageNet-1K, with its weights frozen during training, as detailed in Methods. Specifically, the content loss is computed from the fourth convolutional layer, which captures high-level structural information, whereas the style loss is computed from the Gram matrices of the first five convolutional layers, which capture multi-scale texture statistics. The total loss is then back-propagated through the optoelectronic hybrid neural network to update both the digital encoder and the metasurface phase profile.

Figure 4c shows the evolution of the content and style losses during training. At the early stage, the style loss decreases rapidly while the content loss increases slightly, indicating that the generated images quickly incorporate style features at the expense of partial content distortion. After approximately 100 epochs, the content loss begins to decrease, suggesting that joint optimization gradually balances structural preservation and style transfer. After training, each optimized metasurface functions as an optical style-transfer decoder for the corresponding art style. Representative results are shown in Fig. 4d, including art styles of *The scream*, *Landscape from Saint-Rémy*, and *The magpie*.

In stage S3, the Solver verifies the generated results and compiles the final technical reports, as detailed in Supplementary Note 15. In stage S4, the Verifier cross-checks the reasoning trajectory, generated code, dataset, optimization results, and technical reports. This process identifies three substantive errors. First, the red and blue frequencies are interchanged in the generated technical reports. Second, the number of parameters in the digital encoder is overstated as approximately 1.4 million, whereas the actual value is approximately 398 thousand. Third, the loss calculation is incorrectly described as being performed directly on the optical output; in fact, the perceptual losses are computed from VGG19 feature outputs. The Solver corrects these errors sequentially, with the detected errors and corrections summarized in Fig. 4e. Details of the verification-and-correction loop are provided in Supplementary Notes 16-17.

These demonstrations show that MetaDesigner is not limited to a fixed optimization template, but can automate RGB metalens design, multi-plane full-color CGH, and optical generative modeling. This breadth indicates that the central contribution of MetaDesigner lies not in any single device design, but in its ability to transform natural-language objectives into executable, verifiable, and extensible design workflows.

**Discussion**

MetaDesigner shows that metasurface design does not have to stop at predefined inverse-design pipelines. Starting from a natural-language optical objective, it can plan a design route, gather relevant knowledge, call simulation and optimization tools, generate missing tool code, evaluate the results and prepare technical reports. The three demonstrations — RGB metalens design, multi-plane full-color CGH and optical generative modeling — show that its value is not in optimizing one particular device, but in carrying out a complete optical design process across different tasks. The Verifier is an important part of this process. By checking the reasoning trajectory, generated code, numerical results and reports, the language-model system catches and corrects errors in frequency mapping, numerical aperture estimation, network-parameter counting and loss-function description. This shows that MetaDesigner is not simply producing plausible answers, but can also check parts of the calculation chain that would otherwise require careful human inspection.

Looking ahead, systems like MetaDesigner may become a useful interface between human design intentions and increasingly complex photonic platforms. As metasurfaces move from single-function wavefront elements to multi-functional devices, optical computing systems and generative optical components, future design tasks will require more than running an optimizer on a fixed objective. They will require choosing models, assembling tools, revising assumptions and checking whether the result remains physically meaningful. Extending this approach to broader photonic, electromagnetic or wave-based systems will also require closer links to high-fidelity simulation, experimental feedback, fabrication constraints and field-specific verification standards. In this sense, MetaDesigner is not a final answer to autonomous optical design, but a step toward design agents that can participate in the formulation, execution and validation of complex physical design problems.

## Methods

### Plan-and-Solve prompting

MetaDesigner is built on Plan-and-Solve prompting. For complex tasks that cannot be solved in limited reasoning steps, the agents first generate an executable plan that decomposes the task into simpler sub-tasks. These sub-tasks are then executed sequentially or assigned to sub-agents. During execution, the agents track global progress, consolidate intermediate feedback, and update the plan. Once a sub-task is completed, it is marked as finished in the plan. After all sub-tasks have been completed, the plan is deleted.

### RGB-D dataset and image sources

The RGB-D dataset is derived from the publicly available DeepFocus dataset. The original RGB-D data contains continuous depth information. The original OpenEXR-format images are converted into tensor-format data that can be directly processed by the Optimizer using the function shown in Algorithm 1.

For the optical generative modeling, four paintings are used as content images: *Rainbow over Veere* by Théo van Rysselberghe, *Starry Night Over the Rhône*, *The Starry Night*, and *Sower at Sunset* by Vincent van Gogh. Three paintings are used as style images: *The Scream* by Edvard Munch, *Landscape from Saint-Rémy* by Vincent van Gogh, and *The Magpie* by Claude Monet. These images are shown in Supplementary Fig. 4.

**RAG database and autonomous knowledge retrieval**

In the optical generative model, each generated image is optimized to preserve the structural features of the given content image and the texture features of the given style image. To quantify the performance of the images, the Programmer autonomously generates code for implementing a VGG19-based perceptual loss function. To improve the accuracy of the code, we construct a RAG database from webpages related to VGG19-based image style transfer, as shown in Algorithm 2. MetaDesigner retrieves relevant implementation knowledge from the database and adapts it for our optoelectronic hybrid neural networks.

Beyond manually constructed RAG databases, MetaDesigner is equipped with arXiv and TavilyClient tools for autonomous literature retrieval and online information search, as shown in Algorithms 3 and 4. These tools allow the language-model system to acquire task-relevant external knowledge dynamically, rather than relying solely on knowledge encoded in the pretrained LLM or on retrieval databases provided by experts.

**Incident light information**

The target intensity distributions in the RGB metalens design and the CGH are both static, so the metasurface is illuminated by normally incident RGB plane waves. By contrast, for the optical generative modeling, the generated image should change according to the input content images. Since the metasurface is passive with a fixed phase profile after training, we introduce the dynamic degree of freedom through the amplitude profile of the incident light. In practice, the spatially varying amplitude profiles can be implemented using active devices such as a spatial light modulator (SLM).

In this work, MetaDesigner constructs a digital encoder to map each input content image into the RGB amplitude profiles of the incident optical field. The network architecture of the digital encoder is shown in Supplementary Fig. 5. Each content image is represented as a [3, 512, 512] tensor and is compressed by the digital encoder into a [3, 32, 32] tensor, corresponding to the amplitude profiles of the red, green, and blue incident light.

**Algorithm 1| Holographic image processing**

```
current_dir = os.path.dirname(os.path.abspath(__file__))
def func(distance_min, distance_max):
    """
    Input:
        distance_min: The shortest focal length from the holography to the metasurface, in micrometer.
        distance_max: The longest focal length from the holography to the metasurface, in micrometer.
    """
    file = OpenEXR.InputFile(current_dir+"/image.exr")
    pt = Imath.PixelType(Imath.PixelType.FLOAT)
    dw = file.header()['dataWindow']
    size = (dw.max.x - dw.min.x + 1, dw.max.y - dw.min.y + 1)
    imageG = [Image.frombytes("F", size, file.channel(c, pt)) for c in "G"]
    imageB = [Image.frombytes("F", size, file.channel(c, pt)) for c in "B"]
    imageR = [Image.frombytes("F", size, file.channel(c, pt)) for c in "R"]
    image = np.concatenate((imageR, imageB, imageB),0).transpose(1, 2, 0)
    plt.imshow(image)
    plt.axis("off")
    plt.savefig(current_dir + f"/pic.png", dpi=500, bbox_inches='tight', pad_inches=0)
    plt.show()
    file = OpenEXR.InputFile(current_dir+"/depth.exr")
    pt = Imath.PixelType(Imath.PixelType.FLOAT)
    dw = file.header()['dataWindow']
    size = (dw.max.x - dw.min.x + 1, dw.max.y - dw.min.y + 1)
    depth = np.array([Image.frombytes("F", size, file.channel(c, pt)) for c in "G"])[0]
    depth = np.where(depth > 7, 7, depth)
    depth = depth/depth.max()*(distance_max-distance_min)
    discrete_depth = np.round(depth) + distance_min
    plt.imshow(discrete_depth,'gray')
    plt.axis("off")
    plt.colorbar()
    plt.savefig(current_dir + f"/depth.png", dpi=500, bbox_inches='tight', pad_inches=0)
    plt.show()
    trainData = torch.Tensor([])
    distances = torch.Tensor([])
    masks = torch.Tensor([])
    for depth in np.unique(discrete_depth):
        bool_mask = (discrete_depth == depth)
        bool_mask = np.repeat(bool_mask[:,:, np.newaxis], 3, axis=-1)
        img_mask = image*bool_mask
        img_mask =
img_mask[::int(img_mask.shape[0]/kwargs["Meta_N"]), ::int(img_mask.shape[0]/kwargs["Meta_N"])]
        bool_mask =
```

```
bool_mask[::int(bool_mask.shape[0]/kwargs["Meta_N"]), ::int(bool_mask.shape[0]/kwargs["Meta_N"
])]
        plt.imshow(img_mask)
        plt.axis("off")
        plt.savefig(current_dir + f"/HoloPart{int(depth)}.png", dpi=500, bbox_inches='tight',
pad_inches=0)
        plt.show()
        trainData = torch.cat((trainData, torch.Tensor(img_mask.transpose(2, 0, 1)).unsqueeze(0)),0)
        distances = torch.cat((distances, torch.Tensor([[depth]])),0)
        masks = torch.cat((masks, torch.Tensor(bool_mask.transpose(2, 0, 1)).unsqueeze(0)),0)
    torch.save(trainData, current_dir+'/trainData.pt')
    torch.save(distances, current_dir+'/distance.pt')
    torch.save(masks, current_dir+'/masks.pt')
        return f"The dataset has been created. It contains three files:\n1. The target 3D hologram with
dimensions {trainData.shape}. It is saved to the file path {current_dir}\\trainData.pt.\n2. The focal
length array with dimensions {distances.shape}, which represents the distance from each imaging
plane to the metasurface. It is saved to the file path {current_dir}\distance.pt.\n3. The mask with
dimensions {masks.shape}, corresponding to the target 3D hologram. In this mask, regions with
pattern are set to 1, and regions without pattern are set to 0. It is saved to the file path
{current_dir}\masks.pt."
```

**Algorithm 2| Database construction for optical generative model**

```
if os.path.exists('./chroma_rag_db'):
    shutil.rmtree('./chroma_rag_db')
page_url = "https://blog.csdn.net/2401_88440984/article/details/147496729"
bs4_strainer= bs4.SoupStrainer()
loader = WebBaseLoader(
    web_path=(page_url,),
    bs_kwargs={"parse_only":bs4_strainer}
)
docs = loader.load()
text_splitter = RecursiveCharacterTextSplitter(
    chunk_size=1000,
    chunk_overlap=200,
    add_start_index=True
)
all_splits = text_splitter.split_documents(docs)
embedding = OllamaEmbeddings(model="nomic-embed-text")
vector_store=Chroma(
    collection_name="rag_collection",
    embedding_function=embedding,
    persist_directory="./chroma_rag_db"
)
ids = vector_store.add_documents(documents=all_splits)
```

**Algorithm 3| The arXiv tool for autonomous literature retrieval**

```
from langchain_community.agent_toolkits.load_tools import load_tools
```

```
search_paper = load_tools(["arxiv"])
```

**Algorithm 4| The TavilyClient tool for online information search**

```
from tavily import TavilyClient
from langchain.tools import tool
tavilyClient = TavilyClient()
@tool
def internet_search(
    query: str,
    max_results: int = 5,
    topic: Literal["general", "news", "finance"] = "general",
    include_raw_content: bool = False,
):
  """Run a web search"""
  return tavilyClient.search(
    query,
    max_results=max_results,
    include_raw_content=include_raw_content,
    topic=topic)
```

**Acknowledgments**
Y.Y. discloses support for the research of this work from the Key Research and Development Program of the Ministry of Science and Technology [grant numbers 2022YFA1405200 and 2022YFA1404900], the Fundamental Research Funds for the Zhejiang Provincial Universities [grant number 226-2025-00231], the Science Challenge Project [grant number TZ2025015] and the National Natural Science Foundation of China [grant number U25D8017]. H.C. discloses support for the research of this work from the Key Research and Development Program of the Ministry of Science and Technology [grant number 2022YFA1404704].

**Author Contributions Statement**
Y.Y. and B.W. conceived the idea of this research. B.W. proposed the algorithm structure and performed the simulation. B. X., H. L., and Y. L. assisted in the design of the metasurface. L. Z. and Q. C. assisted in the construction of the LLM agent. Y.Y. and B.W. wrote the paper. All authors shared their insights and contributed to discussions on the results. H.C. and Y.Y. supervised the project.

**Competing Interests Statement**
The authors declare no competing interests.

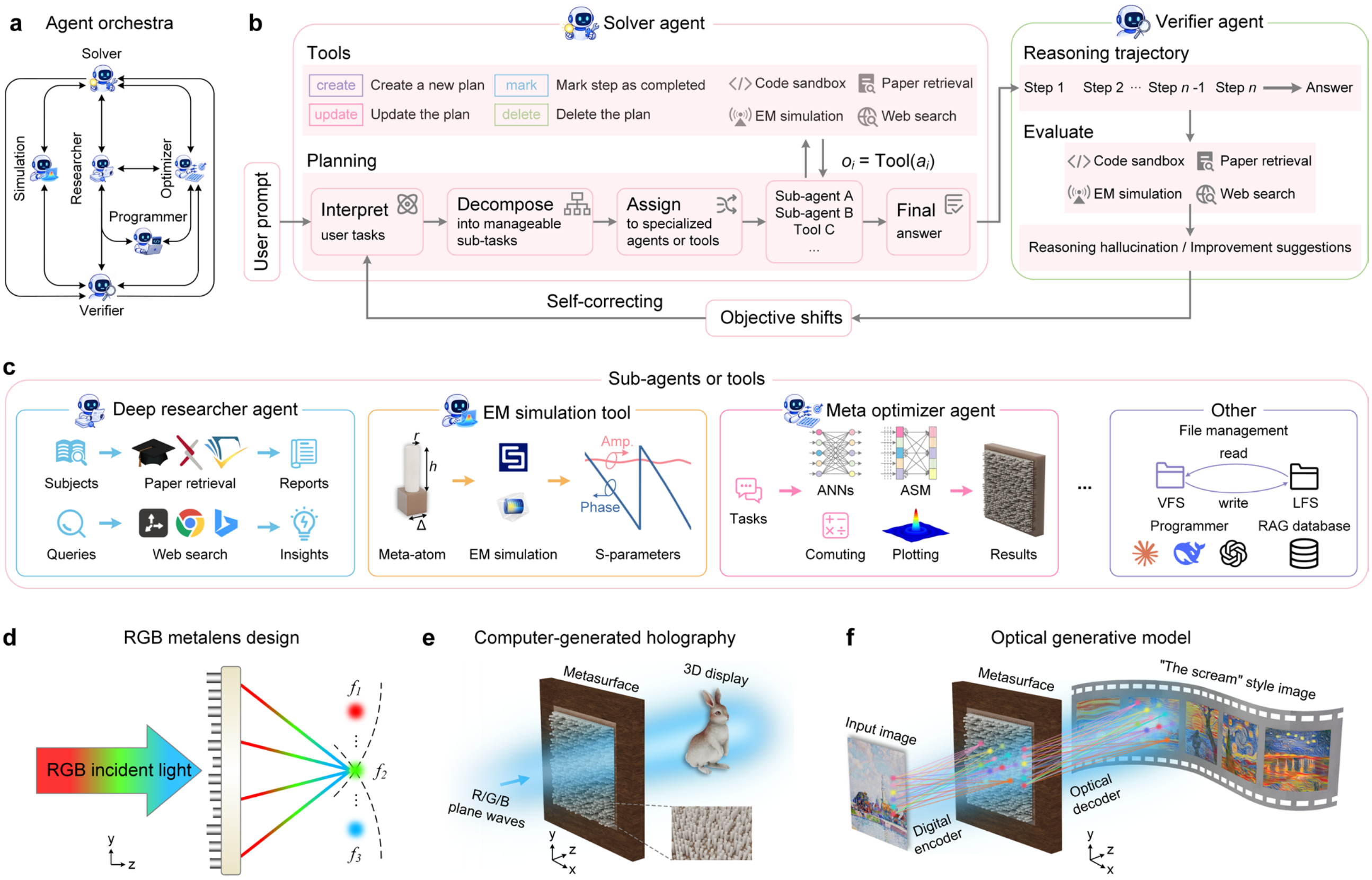

**Fig. 1 | Framework and applications. a**, Agent orchestra showing autonomous interaction among the Solver, Verifier, Researcher, Optimizer, Programmer, and the CST-based electromagnetic simulation tool. **b**, Framework of MetaDesigner. The Solver interprets user prompts, decomposes the design objectives into executable sub-tasks, and orchestrates specialized sub-agents to solve them collaboratively. The Verifier examines the reasoning trajectory, flags potential errors, and provides corrective feedback that the Solver adopts to refine its reasoning. **c**, Specialized capabilities of sub-agents and tools. **d**, RGB metalens design for multi-wavelength point focusing. **e**, CGH for multi-plane image reconstruction. **f**, Optical generative modeling for image style transfer.

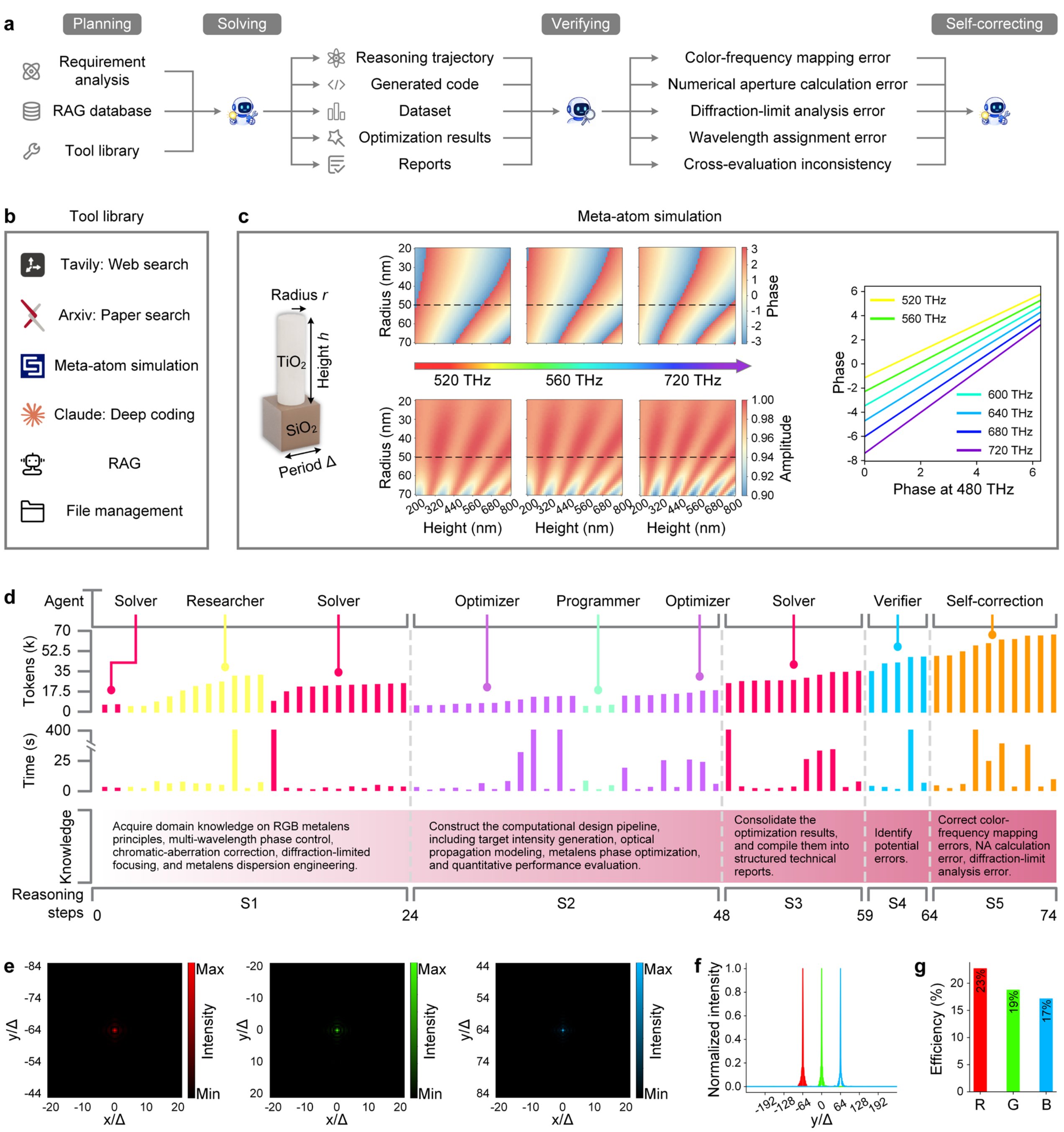

**Fig. 2 | Autonomous RGB metalens designed by MetaDesigner.** **a**, Agentic workflow for RGB metalens design. **b**, Available tools in the library. **c**, Simulation of meta-atom scattering responses with CST. The meta-atom is designed as a $TiO_2$ nanopillar on a $SiO_2$ substrate, with the unit-cell period and nanopillar radius fixed at 160 nm and 50 nm, respectively. **d**, Time and token consumption per reasoning step. The reasoning trajectory consists of 74 steps, taking 16 min 14 s and consuming 1.89 million tokens. **e**, Optical field distributions of red, green, and blue light on the focal plane. **f**, Normalized intensity profiles along the *y*-axis. **g**, The focusing efficiency of red, green, and blue light.

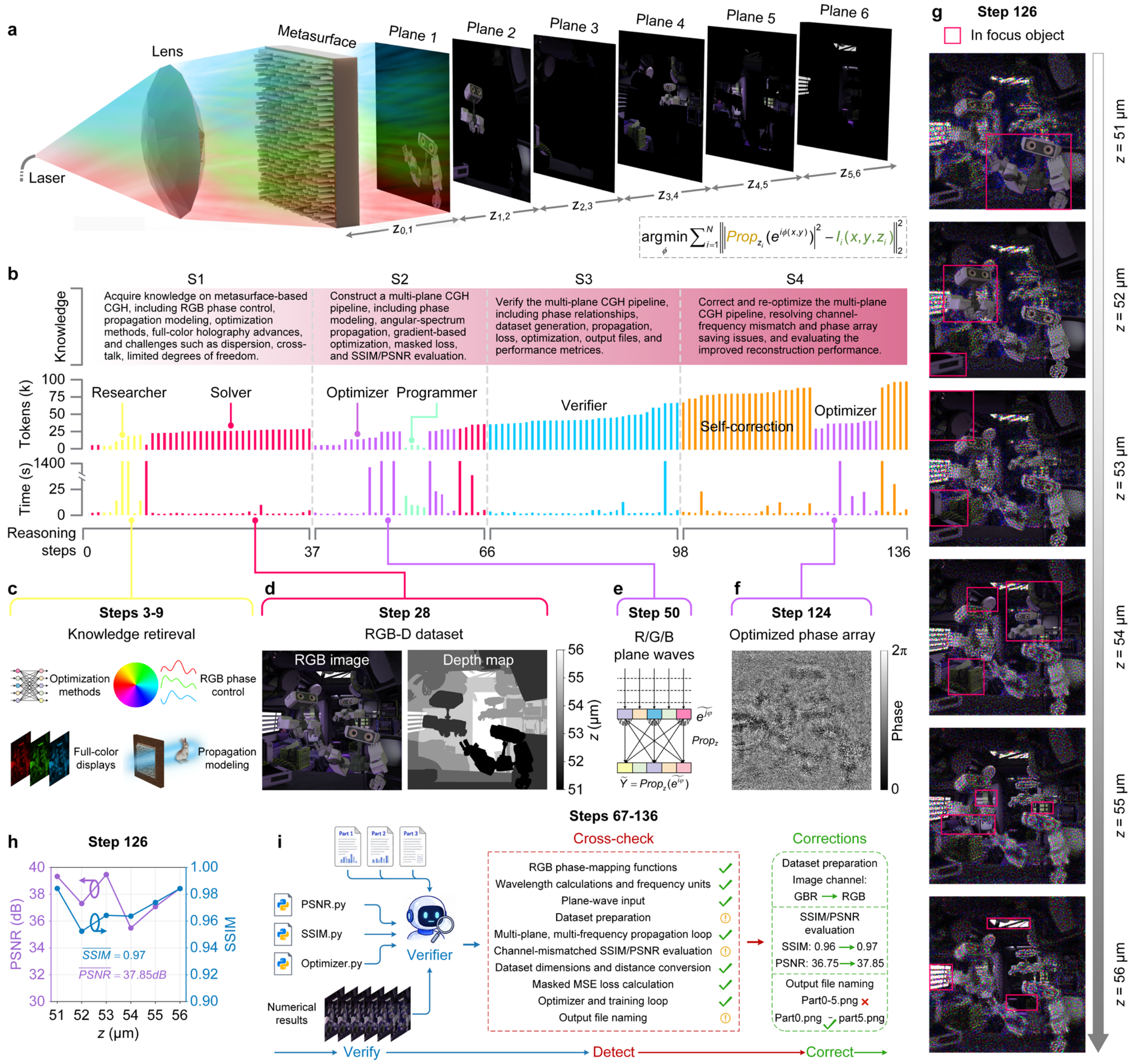

**Fig. 3 | Multi-plane CGH designed by MetaDesigner.** **a**, Schematic of multi-plane CGH. The normally incident red, green, and blue light is modulated by the metasurface to reconstruct distinct full-color holographic images at six image planes. **b**, Time and token consumption per reasoning step. The reasoning trajectory consists of 136 steps, taking 47 min 27 s and consuming 5.62 million tokens. **c**, Schematic of retrieved knowledge, including optimization methods for CGH, RGB phase control, full-color displays, and optical propagation modeling. **d**, Input RGB image and depth map used to generate the target RGB intensity distributions. **e**, Optical propagation modeling based on the ASM. **f**, Optimized phase profile at 480 THz. **g**, Reconstructed holographic images at six image planes, $z_i$ = 51, 52, 53, 54, 55, and 56 μm. **h**, Image fidelity evaluated using PSNR and SSIM, with average values of 37.85 dB and 0.97, respectively. **i**, Detected error and correction details.

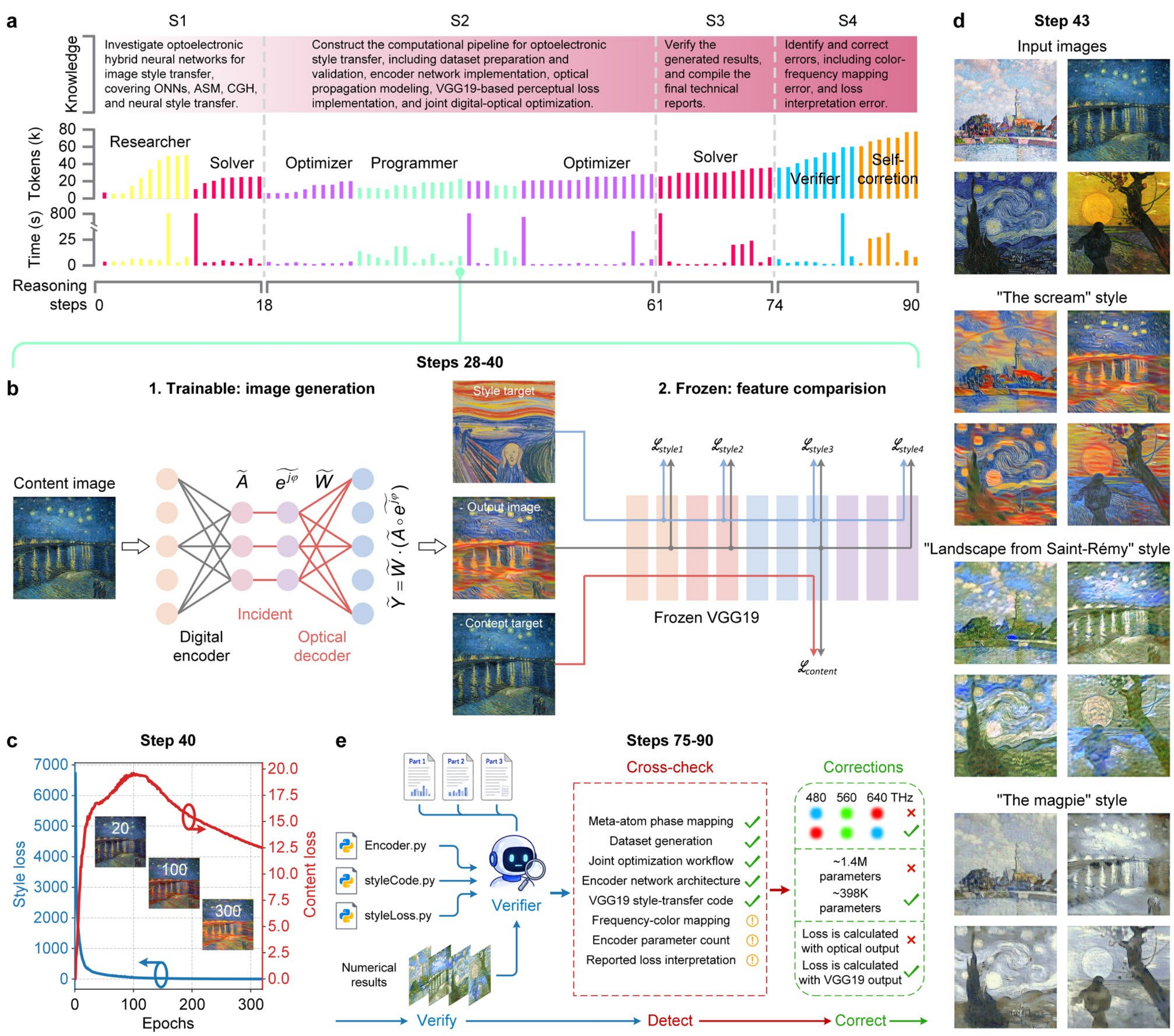

**Fig. 4 | Optical generative modeling for image style transfer designed by MetaDesigner. a**, Time and token consumption per reasoning step. The reasoning trajectory consists of 90 steps, taking 21 min 57 s and consuming 2.53 million tokens. **b**, Training framework of the optoelectronic hybrid neural network, which consists of a shallow digital encoder and a phase-modulated metasurface decoder. The frozen VGG19 is used to implement the perceptual loss function. **c**, Evolution of the content and style losses during training. Insets show the generated images at epochs 20, 100, and 300, illustrating the gradual incorporation of the style feature while preserving the content structure. **d**, Representative style-transferred images for different art styles, including *The scream*, *Landscape from Saint-Rémy*, and *The magpie*. **e**, Detected errors and correction details.